\documentclass[MSNbibl,nameyear,seceqn,dvips]{arxstspdf}
\usepackage{flushend}
\usepackage{stfloats}

\volume{29}
\issue{2}
\pubyear{2014}
\firstpage{254}
\lastpage{258}
\doi{10.1214/14-STS474} 
\referstodoi{10.1214/13-STS457}

\makeatletter

\newcommand{\llvert}{\vert}
\newcommand{\GFI}{Generalized Fiducial Inference }
\newcommand{\GFD}{Generalized Fiducial Distribution }
\newcommand{\bX}{\mathbf{X}}
\newcommand{\bY}{\mathbf{Y}}
\newcommand{\bA}{\mathbf{A}}
\newcommand{\bS}{\mathbf{S}}
\newcommand{\bU}{\mathbf{U}}
\newcommand{\bG}{\mathbf{G}}
\newcommand{\bi}{\mathbf{i}}
\newcommand{\eq}[1]{(\ref{eq:#1})}
\newcommand{\argmin}{\operatorname{\arg\min}}
\renewcommand{\epsilon}{\varepsilon}

\renewcommand{\citep}[1]{(\citeauthor{#1}, \citeyear{#1})}

\makeatother

\begin{document}
\begin{frontmatter}

\vspace*{6pt}\title{Discussion of ``On the Birnbaum Argument for the Strong Likelihood Principle''}
\runtitle{Discussion}

\begin{aug}
\author[a]{\fnms{Jan}~\snm{Hannig}\corref{}\ead[label=e1]{jan.hannig@unc.edu}}
\runauthor{J. Hanning}

\affiliation{University of North Carolina at Chapel Hill}

\address[a]{Jan Hannig is Professor,
Department of Statistics and Operations Research,
University of North Carolina at Chapel Hill, 330 Hanes Hall,
Chapel Hill, North Carolina 27599-3260,
USA \printead{e1}.}
\end{aug}

\begin{abstract}
In this discussion we demonstrate that fiducial distributions provide a natural example of an inference
paradigm that does not obey Strong Likelihood Principle while still satisfying the Weak Conditionality
Principle.
\end{abstract}

%
\end{frontmatter}

Professor Mayo should be congratulated on bringing new light into the
veritable arguments about statistical foundations. It is well
documented that p-values, confidence intervals and hypotheses tests do
not satisfy the Strong Likelihood Principle (SLP). In the next section
we will demonstrate that fiducial distributions provide a natural
example of an inference paradigm that breaks SLP while still satisfying
the Weak Conditionality Principle (WCP).

\section{History of Fiducial Inference}

The origin of \GFI can be traced back to R. A. Fisher
(\citeauthor{Fisher1930}, \citeyear{Fisher1930}, \citeyear{Fisher1933}, \citeyear{Fisher1935a}) who introduced the concept
of a fiducial distribution for a parameter and proposed the
use of this fiducial distribution in place of the Bayesian posterior
distribution.
In the case of a one-parameter family of distributions, Fisher gave the
following
definition for a fiducial density $r(\theta)$ of the parameter based on
a single observation $x$ for the case where the cdf $F(x,\theta)$ is a
decreasing function of $\theta$:
%
\begin{equation}
\label{eq:FisherFiducial} r(\theta) = -\frac{\partial F(x,\theta)}{\partial\theta}.
\end{equation}
%
For multiparameter families of distributions Fisher did not give a
formal definition.
Moreover, the fiducial approach led to confidence sets whose
frequentist coverage
probabilities were close to the claimed confidence levels but they
were not exact in the frequentist sense. Fisher's proposal led to
major discussions among the prominent statisticians of the 1930s,
40s and 50s (e.g., \citeauthor{Dempster1966}, \citeyear{Dempster1966}, \citeyear{Dempster1968};
\citeauthor{Fraser1961b}, \citeyear{Fraser1961a},
\citeyear{Fraser1961b}, \citeyear{Fraser1966}, \citeyear{Fraser1968};
 \cite{Jeffreys1940}; \cite{Lindley1958};
\cite{Stevens1950}). Many of these discussions focused on the
nonexactness of the confidence sets and also on the nonuniqueness of fiducial
distributions. The latter part of the 20th century has seen only a
handful of publications (\citeauthor{Barnard1995}, \citeyear{Barnard1995}; \cite{DawidStoneZidek1973};
\cite{Salome1998}; \cite{DawidStone1982}; \cite{Wilkinson1977}) as the fiducial approach
fell into disfavor and
became a topic of historical interest only.

Since the mid-2000s, there has been a true resurrection of interest in
modern modifications of fiducial inference. These approaches have
become known under the umbrella name of \emph{distributional
inference}. This increase of interest came both in the number of
different approaches to the problem and the number of researchers
working on these problems, and manifested itself in an increasing
number of publications in premier journals. The common thread for these
approaches is a definition of inferentially meaningful probability
statements about subsets of the parameter space without the need for
subjective prior information.

These modern approaches include the Dempster--Shafer theory (\cite
{Dempster2008}; \cite{EdlefsenLiuDempster2009}) and its recent
extension called \textit{inferential models} (\cite
{MartinZhangLiu2010}; \cite{ZhangLiu2011}; \cite{MartinLiu2013b}, \citeyear{MartinLiu2013a},
\citeyear{MartinLiu2013c}, \citeyear{MartinLiu2013d}). A somewhat different approach termed
\textit{confidence distributions} looks at the problem of
obtaining an inferentially meaningful distribution on the parameter
space from a purely frequentist point of view \citep{XieSingh2013}.
One of the main contributions of this approach is the ability to
combine information from disparate sources with deep implications for
meta analysis (\cite{SchwederHjort2002}; \cite{SinghXieStrawderman2005};
\cite{XieSinghStrawderman2011}; \cite{HannigXie2012}; \cite{XieEtAl2013}). Another more
mature approach is called \emph{objective Bayesian inference} that
aims at finding nonsubjective model-based priors. An example of a
recent breakthrough in this area is the modern development of reference
priors (\cite{Berger1992}; \cite{BergerSun2008}; \cite{BergerBernardoSun2009};
\citeyear{BergerBernardoSun2012}; \cite{BayarriEtAl2012}). Another related approach is
based on higher order likelihood expansions and implied data dependent priors
(\cite{FraserFraserStaicu2010};
\cite{Fraser2004},
\citeyear{Fraser2011};
\cite{Fraser:2008tu};
\cite{FraserReidMarrasYi2010};
\cite{Fraser:2005tc}).
There is also important initial work showing how some simple fiducial
distributions that are not Bayesian posteriors naturally arise within
the decision theoretical framework \citep{TaraldsenLindqvist2013}.

Arguably, \GFI has been on the forefront of the modern fiducial
revival. Starting in the early 1990s, the work of Tsui and Weerahandi (\citeyear{TsuiWeerahandi1989},
\citeyear{TsuiWeerahandi1991}) and \citeauthor{Weerahandi1993} (\citeyear{Weerahandi1993},
\citeyear{Weerahandi1994}, \citeyear{Weerahandi1995}) on \emph{generalized confidence intervals}
and the work of \citet{Chiang2001} on the \textit{surrogate variable method}
for obtaining confidence intervals for variance components led to the
realization that there was a connection between these new
procedures and fiducial inference. This realization evolved through a
series of works in the early 2000s (\cite{Hannig2009}; \cite{HannigIyerPatterson2006};
\cite{IyerWangMathew2004}; \cite{PattersonHannigIyer2004}).
The strengths and limitations of the fiducial approach are starting to be
better understood; see especially \citeauthor{Hannig2009} (\citeyear{Hannig2009}, \citeyear{Hannig2013}). In particular,
the asymptotic exactness of fiducial confidence sets, under fairly
general conditions, was established in \citet{Hannig2013};
\citet{HannigIyerPatterson2006}; \citet{SondereggerHannig2013}.
Generalized fiducial inference has also been extended to prediction
problems in \citet{WangHannigIyer2012}.
Computational issues were discussed in \citet{CisewskiHannig2012},
\citet{HannigLaiLee2013}, and model selection in the context of \GFI has been
studied in \citet{HannigLee2009}; \citet{LaiHannigLee2013}.

\section{Generalized Fiducial Distribution and the Weak Conditionality
Principle}

Most modern incarnations of fiducial inference begin with expressing
the relationship between the
data, $\bX$, and the parameters, $\xi$, as
%
\begin{equation}
\label{eq:StructuralEq} \bX= \bG(\bU,\xi),
\end{equation}
where $\bG(\cdot,\cdot)$ is termed the \emph{data generating
equation} (also called the association equation or structural equation)
and $\bU$ is the random component of this data generating equation
whose distribution is free of parameters and completely known.

After observing the data $\mathbf{x}$ the next step is to
use the known distribution of $\bU$ and the inverse of the data \eq
{StructuralEq} to define probabilities for the subsets of the parameter
space. 
In particular, \GFI defines a distribution on the parameter space
as the weak limit as $\epsilon\to0$ of the conditional distribution
%
\begin{equation}
\label{eq:FiducialLimit}\hspace*{2pt} {\mathop{\argmin}_{\xi} \bigl\|{\mathbf{x}}-\bG\bigl(\bU^\star,
\xi\bigr)\bigr\| \bigm|\Bigl\{\min_\xi\bigl\|{\mathbf{x}}-\bG\bigl(
\bU^\star,\xi\bigr)\bigr\|\leq \epsilon}\Bigr\},
\end{equation}
where $\bU^\star$ has the same distribution as $\bU$. If there are
multiple values minimizing the norm, the operator $\argmin_\xi$
selects one of them (possibly at random). We stress at this point that
the \GFD is not unique. For example, different data generating
equations can give a somewhat different Generalized Fiducial Distribution. Notice also that if
$P(\min_\xi\|{\mathbf{x}}-\bG(\bU^\star,\xi)\|=0)>0$,
which is the case for discrete distributions, the limit in \eq
{FiducialLimit} is the conditional distribution evaluated at $\epsilon=0$.

The conditional form of \eq{FiducialLimit} immediately implies the
Weak Conditional Principle for the limiting Generalized Fiducial Distribution. To demonstrate this,
let us consider the two-instrument example of (\cite{Cox1958}) (see also
Section~4.1 of the discussed article). The data generating equation can
be written in a hierarchical form:
\begin{eqnarray*}
M&=&1+I_{(0,1/2)}(U),
\\
X&=&\theta+\sigma_MZ,
\end{eqnarray*}
where $U\sim U(0,1)$ and $Z\sim N(0,1)$ are independent and the
precisions $\sigma_1<<\sigma_2$ are known. If both $X=x$ the
measurement made and $M=m$ the instrument used ($m=1,2$ for machine 1
and 2 respectively) are observed, the conditional distribution \eq
{FiducialLimit} is $N(x,\sigma_m^2)$, only taking into account the
experiment actually performed. On the other hand, if only $M$ is
unobserved, then the conditional distribution \eq{FiducialLimit} is the
mixture $0.5 N(\theta,\sigma_1^2)+0.5 N(\theta,\sigma_2^2)$. As
claimed, the \GFD follows WCP in this example.

\section{Generalized Fiducial Distribution and the Strong Likelihood
Principle}

In general, the \GFD does not satisfy the Strong Likelihood principle.
We first demonstrate this on inference for geometric distribution. To
begin, we perform some preliminary calculations.
Let $X$ be a random variable with discrete distribution function
$F(x,\xi)$. Let us assume for simplicity of presentation that for each
fixed $x$, $F(x,\xi)$ is monotone in $\xi$ and spans the whole $[0,1]$.
The inverse distribution function $F^{-1}(u,\xi)=\inf\{x\dvtx  F(x,\xi
)\geq u\}$ forms a natural data generating equation
\[
X=F^{-1}(U,\xi), \quad U\sim(0,1).
\]
The minimizer in \eq{FiducialLimit} is not unique, but any fiducial
distribution will have a distribution function satisfying $1-F(x,\xi
)\leq H(\xi)\leq1-F(x_-,\xi)$ if $F(x,\xi)$ is decreasing in $\xi$
and $F(x_-,\xi)\leq H(\xi)\leq F(x,\xi)$ if $F(x,\xi)$ is
increasing. To resolve this nonuniqueness, \citet{Hannig2009} and \citet
{Efron1998} recommend using the half correction which is the mixture
distribution with distribution functions $H(\xi)=1-(F(x,\xi
)+F(x_-,\xi))/2$ if $F(x,\xi)$ is decreasing in $\xi$ or $H(\xi
)=(F(x,\xi)+F(x_-,\xi))/2$ if $F(x,\xi)$ increasing.

Let us now consider observing a random variable $N=n$ following the
Geometric$(p)$ distribution. SLP implies that the inference based on
observing $N=n$ should be the same as inference based on observing
$X=1$ where $X$ is Binomial$(n,p)$. However, the Geometric based \GFD
has a distribution function between\vadjust{\goodbreak} $1-(1-p)^{n-1}\leq H_G(p)\leq
1-(1-p)^{n}$. The binomial based \GFD uses bounds
$1-(1-p)^{n}-np(1-p)^{n-1} \leq H_B(p)\leq1-(1-p)^{n}$. Thus, the
effect of the stopping rule demonstrates itself in the \GFI through the
lower bound that is much closer to the upper bound in the case of
geometric distribution. (We remark that one cannot ignore the lower
bound, as the upper bound is used to form upper confidence intervals
and the lower bound is used for lower confidence intervals on $p$.) To
conclude, the fiducial distribution in this example depends on both the
distribution function of $x$ and also on the distribution function of $x-1$.

Let us now turn our attention to continuous distributions. In
particular, assume that the parameter $\xi\in\Theta\subset\mathbb
R^p$ is $p$-dimensional and that the inverse to \eq{StructuralEq} $\bG
^{-1}({\mathbf{x}},\xi) = {\mathbf{u}}$ exists.
Then under some differentiability assumptions, \citet{Hannig2013} shows
that the generalized fiducial distribution is absolutely continuous
with density
%
\begin{equation}
\label{eq:GreatFiducial} r(\xi)=\frac{f({\mathbf{x}},\xi) J({\mathbf{x}},\xi)}{\int_\Theta f({\mathbf{x}},\xi') J({\mathbf{x}},\xi') \,d\xi'},
\end{equation}
where $f({\mathbf{x}},\xi)$ is the likelihood and the
function $J({\mathbf{x}},\xi)$ is
%
\begin{eqnarray}
\label{eq:RecommendedJacobian2}
&&J(\mathbf{x},\xi)\nonumber\hspace*{-20pt}\\[-6pt]\\[-10pt]
&&\quad = \mathop{\sum_{\bi=(i_1,\ldots
,i_p)}}_{ 1\leq i_1<\cdots<i_p\leq n}\biggl|
\det\biggl(\frac
{d}{d\xi} \bG(\mathbf{u},\xi)\Bigl|_{\mathbf{u}=
\bG^{-1}(\mathbf{x},\xi)}\biggr)_\bi \biggr|,\nonumber\hspace*{-20pt}
\end{eqnarray}
where $\frac{d}{d\xi} \bG(\mathbf{u},\xi)$ is the
$n\times p$ Jacobian matrix of partial derivatives computed with
respect of components of $\xi$. The sum in \eq{RecommendedJacobian2}
spans over all $p$-tuples of indexes $\bi=(1\leq i_1<\cdots< i_p\leq
n)\subset\{1, \ldots, n\}$. Additionally, for any $n\times p$ matrix
$J$, the sub-matrix $(J)_\bi$ is the $p\times p$ matrix containing the
rows $\bi=(i_1,\ldots,i_p)$ of $A$. 
The form of \eq{GreatFiducial} suggests that as long as the Jacobian
$J({\mathbf{x}},\xi)$ does not separate into $J({\mathbf{x}},\xi)=f({\mathbf{x}})g(\xi)$, in which
case the \GFD is the same as the Bayes posterior with $g(\xi)$ used as
a prior, the \GFD does not satisfy SLP due to the dependance on ${d}\bG
({\mathbf{u}},\xi)/{d\xi}$.

\section{\GFD and Sufficiency Principle}
Whether the \GFD satisfies the sufficiency principle depends entirely
on what data generating equation is chosen.
For example, let us assume that $\bY=(\bS(\bX),\bA(\bX))'$, where
$\bS$ is a $p$-dimensional sufficient\vadjust{\goodbreak} and $\bA$ is ancillary and $\bX
$ satisfies \eq{StructuralEq}.
Because $d\bA/d\xi=0$, the sum in \eq{RecommendedJacobian2} contains
only one nonzero term:
%
\begin{equation}
\label{eq:ancillary}\quad  J({\mathbf{x}},\xi) = \biggl\llvert \det \biggl(
\frac
{d}{d\xi} \bS\bigl(\bG({\mathbf{u}},\xi)\bigr)\Bigl\vert
_{{\mathbf{u}}=G^{-1}({\mathbf{x}},\xi)} \biggr)\biggr\vert .
\end{equation}

Let ${\mathbf{s}}=\bS({\mathbf{x}})$ and ${\mathbf{a}}=\bA({\mathbf{x}})$ be the observed values
of the sufficient and ancillary statistics respectively.
To interpret the \GFD assume that there is a unique $\xi$ solving
${\mathbf{s}}=\bS(\bG({\mathbf{u}},\xi))$ for
every ${\mathbf{u}}$ and denote this solution $Q_{\mathbf{s}}({\mathbf{u}})=\xi$. Also assume that the
ancillary data generating equation $\bA(\bG({\mathbf{u}},\xi))=\bA({\mathbf{u}})$ is not a function of $\xi
$. A~straightforward calculation shows that the fiducial density \eq
{GreatFiducial} with \eq{ancillary} is the conditional distribution of
$Q_{{\mathbf{s}}}(\bU^\star)\mid\bA(\bU^\star)={\mathbf{a}}$. We conclude that this choice of data generating
equation leads to inference based on sufficient statistics conditional
on the ancillary. However, we still do not expect the SLP to hold in
general even for this data generating equation.

Heuristically, this is because GFI is using not only the data observed,
but also the data that based on the data generating equation could have
been observed in the neighborhood of the observed data.

\section{Final Remarks}

Let us close with discussing the example of Section~3.1. While the
paper is not very clear on the exact specification of the events, it
appears that for experiment 1 we observe the event
\[
O_1=\{ \bar y_{169} > 1.96\sigma/\sqrt{169}\},
\]
while for the experiment 2 we observe
\begin{eqnarray*}
O_2 &=&\{\bar y_k \leq1.96\sigma/\sqrt{k}, k=1,\ldots,
168, \\
&&\hphantom{\{}\bar y_{169} > 1.96\sigma/\sqrt{169}\}.
\end{eqnarray*}
Since $O_2\subset O_1$, we see that the likelihood
\[
P_\theta(O_2)=P_\theta(O_2\mid
O_1)P(O_1).
\]
Consequently, we would have an SLP pair if and only if $P_\theta
(O_2\mid O_1)$ was a constant as a function of $\theta$. However, this
is not the case, as clearly $P_0(O_2\mid O_1)>0$ and $\lim_{\theta\to
\infty} P_\theta(O_2\mid O_1)=0$. Consequently, we do not have an SLP pair.

\section*{Acknowledgments}
Supported in part by the National Science Foundation Grants
1007543 and 1016441.




\begin{thebibliography}{62}

\bibitem[\protect\citeauthoryear{Barnard}{1995}]{Barnard1995}
\begin{barticle}[auto:STB|2014/05/27|07:42:02]
\bauthor{\bsnm{Barnard},~\bfnm{G.~A.}\binits{G.~A.}}
(\byear{1995}).
\btitle{Pivotal models and the fiducial argument}.
\bjournal{Internat. Statist. Rev.}
\bvolume{63}
\bpages{309--323}.
\end{barticle}\vadjust{\goodbreak}
\bptok{imsref}%
\endbibitem

\bibitem[\protect\citeauthoryear{Bayarri et~al.}{2012}]{BayarriEtAl2012}
\begin{barticle}[mr]
\bauthor{\bsnm{Bayarri},~\bfnm{M.~J.}\binits{M.~J.}},
\bauthor{\bsnm{Berger},~\bfnm{J.~O.}\binits{J.~O.}},
\bauthor{\bsnm{Forte},~\bfnm{A.}\binits{A.}} \AND
\bauthor{\bsnm{Garc{\'{\i}}a-Donato},~\bfnm{G.}\binits{G.}}
(\byear{2012}).
\btitle{Criteria for {B}ayesian model choice with application to variable selection}.
\bjournal{Ann. Statist.}
\bvolume{40}
\bpages{1550--1577}.
\bid{doi={10.1214/12-AOS1013}, issn={0090-5364}, mr={3015035}}
\end{barticle}
\bptok{imsref}%
\endbibitem

\bibitem[\protect\citeauthoryear{Berger}{1992}]{Berger1992}
\begin{bincollection}[auto:STB|2014/05/27|07:42:02]
\bauthor{\bsnm{Berger},~\bfnm{James~O.}\binits{J.~O.}} \AND
\bauthor{\bsnm{Bernardo},~\bfnm{J. M.}\binits{J. M.}}
 (\byear{1992}).
 \btitle{On the development of reference priors}. In
\bbooktitle{Bayesian Statistics 4}
(\beditor{\bfnm{J.~M.}\binits{J.~M.}~\bsnm{Bernardo}},
\beditor{\bfnm{J.~O.}\binits{J.~O.}~\bsnm{Berger}},
\beditor{\bfnm{A.~P.}\binits{A.~P.}~\bsnm{Dawid}} \AND
\beditor{\bfnm{A.~F.~M.}\binits{A.~F.~M.}~\bsnm{Smith}}, eds.)
\bpages{35--60}.
\bpublisher{Oxford Univ. Press},
\blocation{New York}.
\bid{mr={1380269}}
\end{bincollection}
\bptok{imsref}%
\endbibitem

\bibitem[\protect\citeauthoryear{Berger, Bernardo and Sun}{2009}]{BergerBernardoSun2009}
\begin{barticle}[mr]
\bauthor{\bsnm{Berger},~\bfnm{James~O.}\binits{J.~O.}},
\bauthor{\bsnm{Bernardo},~\bfnm{Jos{\'e}~M.}\binits{J.~M.}} \AND
\bauthor{\bsnm{Sun},~\bfnm{Dongchu}\binits{D.}}
(\byear{2009}).
\btitle{The formal definition of reference priors}.
\bjournal{Ann. Statist.}
\bvolume{37}
\bpages{905--938}.
\bid{doi={10.1214/07-AOS587}, issn={0090-5364}, mr={2502655}}
\end{barticle}
\bptok{imsref}%
\endbibitem

\bibitem[\protect\citeauthoryear{Berger, Bernardo and Sun}{2012}]{BergerBernardoSun2012}
\begin{barticle}[mr]
\bauthor{\bsnm{Berger},~\bfnm{James~O.}\binits{J.~O.}},
\bauthor{\bsnm{Bernardo},~\bfnm{Jose~M.}\binits{J.~M.}} \AND
\bauthor{\bsnm{Sun},~\bfnm{Dongchu}\binits{D.}}
(\byear{2012}).
\btitle{Objective priors for discrete parameter spaces}.
\bjournal{J. Amer. Statist. Assoc.}
\bvolume{107}
\bpages{636--648}.
\bid{doi={10.1080/01621459.2012.682538}, issn={0162-1459}, mr={2980073}}
\end{barticle}
\bptok{imsref}%
\endbibitem

\bibitem[\protect\citeauthoryear{Berger and Sun}{2008}]{BergerSun2008}
\begin{barticle}[mr]
\bauthor{\bsnm{Berger},~\bfnm{James~O.}\binits{J.~O.}} \AND
\bauthor{\bsnm{Sun},~\bfnm{Dongchu}\binits{D.}}
(\byear{2008}).
\btitle{Objective priors for the bivariate normal model}.
\bjournal{Ann. Statist.}
\bvolume{36}
\bpages{963--982}.
\bid{doi={10.1214/07-AOS501}, issn={0090-5364}, mr={2396821}}
\end{barticle}
\bptok{imsref}%
\endbibitem


\bibitem[\protect\citeauthoryear{Chiang}{2001}]{Chiang2001}
\begin{barticle}[mr]
\bauthor{\bsnm{Chiang},~\bfnm{Andy~K.~L.}\binits{A.~K.~L.}}
(\byear{2001}).
\btitle{A simple general method for constructing confidence intervals for functions of variance components}.
\bjournal{Technometrics}
\bvolume{43}
\bpages{356--367}.
\bid{doi={10.1198/004017001316975943}, issn={0040-1706}, mr={1943189}}
\end{barticle}
\bptok{imsref}%
\endbibitem

\bibitem[\protect\citeauthoryear{Cisewski and Hannig}{2012}]{CisewskiHannig2012}
\begin{barticle}[mr]
\bauthor{\bsnm{Cisewski},~\bfnm{Jessi}\binits{J.}} \AND
\bauthor{\bsnm{Hannig},~\bfnm{Jan}\binits{J.}}
(\byear{2012}).
\btitle{Generalized fiducial inference for normal linear mixed models}.
\bjournal{Ann. Statist.}
\bvolume{40}
\bpages{2102--2127}.
\bid{doi={10.1214/12-AOS1030}, issn={0090-5364}, mr={3059078}}
\end{barticle}
\bptok{imsref}%
\endbibitem

\bibitem[\protect\citeauthoryear{Cox}{1958}]{Cox1958}
\begin{barticle}[mr]
\bauthor{\bsnm{Cox},~\bfnm{D.~R.}\binits{D.~R.}}
(\byear{1958}).
\btitle{Some problems connected with statistical inference}.
\bjournal{Ann. Math. Statist.}
\bvolume{29}
\bpages{357--372}.
\bid{issn={0003-4851}, mr={0094890}}
\end{barticle}
\bptok{imsref}%
\endbibitem

\bibitem[\protect\citeauthoryear{Dawid and Stone}{1982}]{DawidStone1982}
\begin{barticle}[mr]
\bauthor{\bsnm{Dawid},~\bfnm{A.~P.}\binits{A.~P.}} \AND
\bauthor{\bsnm{Stone},~\bfnm{M.}\binits{M.}}
(\byear{1982}).
\btitle{The functional-model basis of fiducial inference}.
\bjournal{Ann. Statist.}
\bvolume{10}
\bpages{1054--1074}.
\bid{issn={0090-5364}, mr={0673643}}
\bptnote{check related}%
\end{barticle}
\bptok{imsref}%
\endbibitem

\bibitem[\protect\citeauthoryear{Dawid, Stone and Zidek}{1973}]{DawidStoneZidek1973}
\begin{barticle}[mr]
\bauthor{\bsnm{Dawid},~\bfnm{A.~P.}\binits{A.~P.}},
\bauthor{\bsnm{Stone},~\bfnm{M.}\binits{M.}} \AND
\bauthor{\bsnm{Zidek},~\bfnm{J.~V.}\binits{J.~V.}}
(\byear{1973}).
\btitle{Marginalization paradoxes in {B}ayesian and structural inference}.
\bjournal{J. R. Stat. Soc. Ser. B Stat. Methodol.}
\bvolume{35}
\bpages{189--233}.
\bid{issn={0035-9246}, mr={0365805}}
\bptnote{check related}%
\end{barticle}
\bptok{imsref}%
\endbibitem

\bibitem[\protect\citeauthoryear{Dempster}{1966}]{Dempster1966}
\begin{barticle}[mr]
\bauthor{\bsnm{Dempster},~\bfnm{A.~P.}\binits{A.~P.}}
(\byear{1966}).
\btitle{New methods for reasoning towards posterior distributions based on sample data}.
\bjournal{Ann. Math. Statist.}
\bvolume{37}
\bpages{355--374}.
\bid{issn={0003-4851}, mr={0187357}}
\end{barticle}
\bptok{imsref}%
\endbibitem

\bibitem[\protect\citeauthoryear{Dempster}{1968}]{Dempster1968}
\begin{barticle}[mr]
\bauthor{\bsnm{Dempster},~\bfnm{A.~P.}\binits{A.~P.}}
(\byear{1968}).
\btitle{A generalization of {B}ayesian inference ({w}ith discussion)}.
\bjournal{J. R. Stat. Soc. Ser. B Stat. Methodol.}
\bvolume{30}
\bpages{205--247}.
\bid{issn={0035-9246}, mr={0238428}}
\end{barticle}
\bptok{imsref}%
\endbibitem

\bibitem[\protect\citeauthoryear{Dempster}{2008}]{Dempster2008}
\begin{barticle}[mr]
\bauthor{\bsnm{Dempster},~\bfnm{A.~P.}\binits{A.~P.}}
(\byear{2008}).
\btitle{The {D}empster--{S}hafer calculus for statisticians}.
\bjournal{Internat. J. Approx. Reason.}
\bvolume{48}
\bpages{365--377}.
\bid{doi={10.1016/j.ijar.2007.03.004}, issn={0888-613X}, mr={2419025}}
\end{barticle}
\bptok{imsref}%
\endbibitem

\bibitem[\protect\citeauthoryear{Edlefsen, Liu and Dempster}{2009}]{EdlefsenLiuDempster2009}
\begin{barticle}[mr]
\bauthor{\bsnm{Edlefsen},~\bfnm{Paul~T.}\binits{P.~T.}},
\bauthor{\bsnm{Liu},~\bfnm{Chuanhai}\binits{C.}} \AND
\bauthor{\bsnm{Dempster},~\bfnm{Arthur~P.}\binits{A.~P.}}
(\byear{2009}).
\btitle{Estimating limits from {P}oisson counting data using {D}empster--{S}hafer analysis}.
\bjournal{Ann. Appl. Stat.}
\bvolume{3}
\bpages{764--790}.
\bid{doi={10.1214/00-AOAS223}, issn={1932-6157}, mr={2750681}}
\end{barticle}
\bptok{imsref}%
\endbibitem

\bibitem[\protect\citeauthoryear{Efron}{1998}]{Efron1998}
\begin{barticle}[mr]
\bauthor{\bsnm{Efron},~\bfnm{Bradley}\binits{B.}}
(\byear{1998}).
\btitle{R. {A}. {F}isher in the 21st century (invited paper presented at the 1996 {R}. {A}. {F}isher {L}ecture)}.
\bjournal{Statist. Sci.}
\bvolume{13}
\bpages{95--122}.
\bid{doi={10.1214/ss/1028905930}, issn={0883-4237}, mr={1647499}}
\bptnote{check related}%
\end{barticle}
\bptok{imsref}%
\endbibitem

\bibitem[\protect\citeauthoryear{Fisher}{1930}]{Fisher1930}
\begin{barticle}[auto:STB|2014/05/27|07:42:02]
\bauthor{\bsnm{Fisher},~\bfnm{R.~A.}\binits{R.~A.}}
(\byear{1930}).
\btitle{Inverse probability}.
\bjournal{Math. Proc. Cambridge Philos. Soc.}
\bvolume{XXVI}
\bpages{528--535}.
\end{barticle}
\bptok{imsref}%
\endbibitem

\bibitem[\protect\citeauthoryear{Fisher}{1933}]{Fisher1933}
\begin{barticle}[auto:STB|2014/05/27|07:42:02]
\bauthor{\bsnm{Fisher},~\bfnm{R.~A.}\binits{R.~A.}}
(\byear{1933}).
\btitle{The concepts of inverse probability and fiducial probability referring to unknown parameters}.
\bjournal{Proc. R. Soc. Lond. Ser. A Math. Phys. Eng. Sci.}
\bvolume{139}
\bpages{343--348}.
\end{barticle}
\bptok{imsref}%
\endbibitem

\bibitem[\protect\citeauthoryear{Fisher}{1935}]{Fisher1935a}
\begin{barticle}[auto:STB|2014/05/27|07:42:02]
\bauthor{\bsnm{Fisher},~\bfnm{R.~A.}\binits{R.~A.}}
(\byear{1935}).
\btitle{The fiducial argument in statistical inference}.
\bjournal{Ann. Eugenics}
\bvolume{VI}
\bpages{91--98}.
\end{barticle}
\bptok{imsref}%
\endbibitem

\bibitem[\protect\citeauthoryear{Fraser}{1961a}]{Fraser1961a}
\begin{barticle}[mr]
\bauthor{\bsnm{Fraser},~\bfnm{D.~A.~S.}\binits{D.~A.~S.}}
(\byear{1961}a).
\btitle{On fiducial inference}.
\bjournal{Ann. Math. Statist.}
\bvolume{32}
\bpages{661--676}.
\bid{issn={0003-4851}, mr={0130755}}
\end{barticle}
\bptok{imsref}%
\endbibitem

\bibitem[\protect\citeauthoryear{Fraser}{1961b}]{Fraser1961b}
\begin{barticle}[mr]
\bauthor{\bsnm{Fraser},~\bfnm{D.~A.~S.}\binits{D.~A.~S.}}
(\byear{1961}b).
\btitle{The fiducial method and invariance}.
\bjournal{Biometrika}
\bvolume{48}
\bpages{261--280}.
\bid{issn={0006-3444}, mr={0133910}}
\end{barticle}
\bptok{imsref}%
\endbibitem

\bibitem[\protect\citeauthoryear{Fraser}{1966}]{Fraser1966}
\begin{barticle}[mr]
\bauthor{\bsnm{Fraser},~\bfnm{D.~A.~S.}\binits{D.~A.~S.}}
(\byear{1966}).
\btitle{Structural probability and a generalization}.
\bjournal{Biometrika}
\bvolume{53}
\bpages{1--9}.
\bid{issn={0006-3444}, mr={0196840}}
\end{barticle}
\bptok{imsref}%
\endbibitem

\bibitem[\protect\citeauthoryear{Fraser}{1968}]{Fraser1968}
\begin{bbook}[mr]
\bauthor{\bsnm{Fraser},~\bfnm{D.~A.~S.}\binits{D.~A.~S.}}
(\byear{1968}).
\btitle{The Structure of Inference}.
\bpublisher{Wiley},
\blocation{New York}.
\bid{mr={0235643}}
\end{bbook}
\bptok{imsref}%
\endbibitem

\bibitem[\protect\citeauthoryear{Fraser}{2004}]{Fraser2004}
\begin{barticle}[mr]
\bauthor{\bsnm{Fraser},~\bfnm{D.~A.~S.}\binits{D.~A.~S.}}
(\byear{2004}).
\btitle{Ancillaries and conditional inference}.
\bjournal{Statist. Sci.}
\bvolume{19}
\bpages{333--369}.
\bid{doi={10.1214/088342304000000323}, issn={0883-4237}, mr={2140544}}
\bptnote{check related}%
\end{barticle}
\bptok{imsref}%
\endbibitem

\bibitem[\protect\citeauthoryear{Fraser}{2011}]{Fraser2011}
\begin{barticle}[mr]
\bauthor{\bsnm{Fraser},~\bfnm{D.~A.~S.}\binits{D.~A.~S.}}
(\byear{2011}).
\btitle{Is {B}ayes posterior just quick and dirty confidence?}
\bjournal{Statist. Sci.}
\bvolume{26}
\bpages{299--316}.
\bid{doi={10.1214/11-STS352}, issn={0883-4237}, mr={2918001}}
\end{barticle}
\bptok{imsref}%
\endbibitem

\bibitem[\protect\citeauthoryear{Fraser, Fraser and Staicu}{2010}]{FraserFraserStaicu2010}
\begin{barticle}[mr]
\bauthor{\bsnm{Fraser},~\bfnm{Ailana~M.}\binits{A.~M.}},
\bauthor{\bsnm{Fraser},~\bfnm{D.~A.~S.}\binits{D.~A.~S.}} \AND
\bauthor{\bsnm{Staicu},~\bfnm{Ana-Maria}\binits{A.-M.}}
(\byear{2010}).
\btitle{Second order ancillary: A differential view from continuity}.
\bjournal{Bernoulli}
\bvolume{16}
\bpages{1208--1223}.
\bid{doi={10.3150/10-BEJ248}, issn={1350-7265}, mr={2759176}}
\bptnote{check year}%
\end{barticle}
\bptok{imsref}%
\endbibitem

\bibitem[\protect\citeauthoryear{Fraser and Naderi}{2008}]{Fraser:2008tu}
\begin{barticle}[auto:STB|2014/05/27|07:42:02]
\bauthor{\bsnm{Fraser},~\bfnm{D.~A.~S.}\binits{D.~A.~S.}} \AND
\bauthor{\bsnm{Naderi},~\bfnm{A.}\binits{A.}}
(\byear{2008}).
\btitle{Exponential models: Approximations for probabilities}.
\bjournal{Biometrika}
\bvolume{94}
\bpages{1--9}.
\end{barticle}
\bptok{imsref}%
\endbibitem

\bibitem[\protect\citeauthoryear{Fraser, Reid and Wong}{2005}]{Fraser:2005tc}
\begin{barticle}[auto:STB|2014/05/27|07:42:02]
\bauthor{\bsnm{Fraser},~\bfnm{D.}\binits{D.}},
\bauthor{\bsnm{Reid},~\bfnm{N.}\binits{N.}} \AND
\bauthor{\bsnm{Wong},~\bfnm{A.}\binits{A.}}
(\byear{2005}).
\btitle{What a model with data says about theta}.
\bjournal{Internat. J. Statist. Sci.}
\bvolume{3}
\bpages{163--178}.
\end{barticle}
\bptok{imsref}%
\endbibitem

\bibitem[\protect\citeauthoryear{Fraser et~al.}{2010}]{FraserReidMarrasYi2010}
\begin{barticle}[mr]
\bauthor{\bsnm{Fraser},~\bfnm{D.~A.~S.}\binits{D.~A.~S.}},
\bauthor{\bsnm{Reid},~\bfnm{N.}\binits{N.}},
\bauthor{\bsnm{Marras},~\bfnm{E.}\binits{E.}} \AND
\bauthor{\bsnm{Yi},~\bfnm{G.~Y.}\binits{G.~Y.}}
(\byear{2010}).
\btitle{Default priors for {B}ayesian and frequentist inference}.
\bjournal{J. R. Stat. Soc. Ser. B Stat. Methodol.}
\bvolume{72}
\bpages{631--654}.
\bid{doi={10.1111/j.1467-9868.2010.00750.x}, issn={1369-7412}, mr={2758239}}
\end{barticle}
\bptok{imsref}%
\endbibitem

\bibitem[\protect\citeauthoryear{Hannig}{2009}]{Hannig2009}
\begin{barticle}[mr]
\bauthor{\bsnm{Hannig},~\bfnm{Jan}\binits{J.}}
(\byear{2009}).
\btitle{On generalized fiducial inference}.
\bjournal{Statist. Sinica}
\bvolume{19}
\bpages{491--544}.
\bid{issn={1017-0405}, mr={2514173}}
\end{barticle}
\bptok{imsref}%
\endbibitem

\bibitem[\protect\citeauthoryear{Hannig}{2013}]{Hannig2013}
\begin{barticle}[mr]
\bauthor{\bsnm{Hannig},~\bfnm{Jan}\binits{J.}}
(\byear{2013}).
\btitle{Generalized fiducial inference via discretization}.
\bjournal{Statist. Sinica}
\bvolume{23}
\bpages{489--514}.
\bid{issn={1017-0405}, mr={3086644}}
\end{barticle}
\bptok{imsref}%
\endbibitem

\bibitem[\protect\citeauthoryear{Hannig, Iyer and Patterson}{2006}]{HannigIyerPatterson2006}
\begin{barticle}[mr]
\bauthor{\bsnm{Hannig},~\bfnm{Jan}\binits{J.}},
\bauthor{\bsnm{Iyer},~\bfnm{Hari}\binits{H.}} \AND
\bauthor{\bsnm{Patterson},~\bfnm{Paul}\binits{P.}}
(\byear{2006}).
\btitle{Fiducial generalized confidence intervals}.
\bjournal{J. Amer. Statist. Assoc.}
\bvolume{101}
\bpages{254--269}.
\bid{doi={10.1198/016214505000000736}, issn={0162-1459}, mr={2268043}}
\end{barticle}
\bptok{imsref}%
\endbibitem

\bibitem[\protect\citeauthoryear{Hannig, Lai and Lee}{2014}]{HannigLaiLee2013}
\begin{barticle}[mr]
\bauthor{\bsnm{Hannig},~\bfnm{Jan}\binits{J.}},
\bauthor{\bsnm{Lai},~\bfnm{Randy~C.~S.}\binits{R.~C.~S.}} \AND
\bauthor{\bsnm{Lee},~\bfnm{Thomas~C.~M.}\binits{T.~C.~M.}}
(\byear{2014}).
\btitle{Computational issues of generalized fiducial inference}.
\bjournal{Comput. Statist. Data Anal.}
\bvolume{71}
\bpages{849--858}.
\bid{doi={10.1016/j.csda.2013.03.003}, issn={0167-9473}, mr={3132011}}
\bptnote{check year}%
\end{barticle}
\bptok{imsref}%
\endbibitem

\bibitem[\protect\citeauthoryear{Hannig and Lee}{2009}]{HannigLee2009}
\begin{barticle}[mr]
\bauthor{\bsnm{Hannig},~\bfnm{Jan}\binits{J.}} \AND
\bauthor{\bsnm{Lee},~\bfnm{Thomas~C.~M.}\binits{T.~C.~M.}}
(\byear{2009}).
\btitle{Generalized fiducial inference for wavelet regression}.
\bjournal{Biometrika}
\bvolume{96}
\bpages{847--860}.
\bid{doi={10.1093/biomet/asp050}, issn={0006-3444}, mr={2767274}}
\end{barticle}
\bptok{imsref}%
\endbibitem

\bibitem[\protect\citeauthoryear{Hannig and Xie}{2012}]{HannigXie2012}
\begin{barticle}[mr]
\bauthor{\bsnm{Hannig},~\bfnm{Jan}\binits{J.}} \AND
\bauthor{\bsnm{Xie},~\bfnm{Min-ge}\binits{M.-g.}}
(\byear{2012}).
\btitle{A note on {D}empster--{S}hafer recombination of confidence distributions}.
\bjournal{Electron. J. Stat.}
\bvolume{6}
\bpages{1943--1966}.
\bid{doi={10.1214/12-EJS734}, issn={1935-7524}, mr={2988470}}
\end{barticle}
\bptok{imsref}%
\endbibitem

\bibitem[\protect\citeauthoryear{Iyer, Wang and Mathew}{2004}]{IyerWangMathew2004}
\begin{barticle}[mr]
\bauthor{\bsnm{Iyer},~\bfnm{Hari~K.}\binits{H.~K.}},
\bauthor{\bsnm{Wang},~\bfnm{C.~M.~Jack}\binits{C.~M.~J.}} \AND
\bauthor{\bsnm{Mathew},~\bfnm{Thomas}\binits{T.}}
(\byear{2004}).
\btitle{Models and confidence intervals for true values in interlaboratory trials}.
\bjournal{J. Amer. Statist. Assoc.}
\bvolume{99}
\bpages{1060--1071}.
\bid{doi={10.1198/016214504000001682}, issn={0162-1459}, mr={2109495}}
\end{barticle}
\bptok{imsref}%
\endbibitem

\bibitem[\protect\citeauthoryear{Jeffreys}{1940}]{Jeffreys1940}
\begin{barticle}[mr]
\bauthor{\bsnm{Jeffreys},~\bfnm{Harold}\binits{H.}}
(\byear{1940}).
\btitle{Note on the {B}ehrens--{F}isher formula}.
\bjournal{Ann. Eugenics}
\bvolume{10}
\bpages{48--51}.
\bid{mr={0002080}}
\end{barticle}
\bptok{imsref}%
\endbibitem

\bibitem[\protect\citeauthoryear{Lai, Hannig and Lee}{2013}]{LaiHannigLee2013}
\begin{bmisc}[auto:STB|2014/05/27|07:42:02]
\bauthor{\bsnm{Lai},~\bfnm{R.~C.~S.}\binits{R.~C.~S.}},
\bauthor{\bsnm{Hannig},~\bfnm{J.}\binits{J.}} \AND
\bauthor{\bsnm{Lee},~\bfnm{T.~C.~M.}\binits{T.~C.~M.}}
(\byear{2013}). \bhowpublished{Generalized fiducial inference for ultra high dimensional regression. Available at \arxivurl{arXiv:1304.7847}}.
\end{bmisc}
\bptok{imsref}%
\endbibitem

\bibitem[\protect\citeauthoryear{Lindley}{1958}]{Lindley1958}
\begin{barticle}[mr]
\bauthor{\bsnm{Lindley},~\bfnm{D.~V.}\binits{D.~V.}}
(\byear{1958}).
\btitle{Fiducial distributions and {B}ayes' theorem}.
\bjournal{J. R. Stat. Soc. Ser. B Stat. Methodol.}
\bvolume{20}
\bpages{102--107}.
\bid{issn={0035-9246}, mr={0095550}}
\end{barticle}
\bptok{imsref}%
\endbibitem


\bibitem[\protect\citeauthoryear{Martin and Liu}{2013a}]{MartinLiu2013b}
\begin{bmisc}[auto:STB|2014/05/27|07:42:02]
\bauthor{\bsnm{Martin},~\bfnm{R.}\binits{R.}} \AND
\bauthor{\bsnm{Liu},~\bfnm{C.}\binits{C.}}
(\byear{2013a}). \bhowpublished{{Conditional inferential models: Combining information for prior-free probabilistic inference}. Preprint}.
\end{bmisc}
\bptok{imsref}%
\endbibitem


\bibitem[\protect\citeauthoryear{Martin and Liu}{2013b}]{MartinLiu2013a}
\begin{barticle}[mr]
\bauthor{\bsnm{Martin},~\bfnm{Ryan}\binits{R.}} \AND
\bauthor{\bsnm{Liu},~\bfnm{Chuanhai}\binits{C.}}
(\byear{2013b}).
\btitle{Inferential {m}odels: {A}~{f}ramework for {p}rior-{f}ree {p}osterior {p}robabilistic {i}nference}.
\bjournal{J. Amer. Statist. Assoc.}
\bvolume{108}
\bpages{301--313}.
\bid{doi={10.1080/01621459.2012.747960}, issn={0162-1459}, mr={3174621}}
\end{barticle}
\bptok{imsref}%
\endbibitem

\bibitem[\protect\citeauthoryear{Martin and Liu}{2013c}]{MartinLiu2013c}
\begin{bmisc}[auto:STB|2014/05/27|07:42:02]
\bauthor{\bsnm{Martin},~\bfnm{R.}\binits{R.}} \AND
\bauthor{\bsnm{Liu},~\bfnm{C.}\binits{C.}}
(\byear{2013c}). \bhowpublished{Marginal inferential models: prior-free probabilistic inference on interest parameters. Preprint}.
\end{bmisc}
\bptok{imsref}%
\endbibitem

\bibitem[\protect\citeauthoryear{Martin and Liu}{2013d}]{MartinLiu2013d}
\begin{bmisc}[auto:STB|2014/05/27|07:42:02]
\bauthor{\bsnm{Martin},~\bfnm{R.}\binits{R.}} \AND
\bauthor{\bsnm{Liu},~\bfnm{C.}\binits{C.}}
(\byear{2013d}). \bhowpublished{{On a 'plausible' interpretation of p-values}. Preprint}.
\end{bmisc}
\bptok{imsref}%
\endbibitem

\bibitem[\protect\citeauthoryear{Martin, Zhang and Liu}{2010}]{MartinZhangLiu2010}
\begin{barticle}[mr]
\bauthor{\bsnm{Martin},~\bfnm{Ryan}\binits{R.}},
\bauthor{\bsnm{Zhang},~\bfnm{Jianchun}\binits{J.}} \AND
\bauthor{\bsnm{Liu},~\bfnm{Chuanhai}\binits{C.}}
(\byear{2010}).
\btitle{Dempster--{S}hafer theory and statistical inference with weak beliefs}.
\bjournal{Statist. Sci.}
\bvolume{25}
\bpages{72--87}.
\bid{doi={10.1214/10-STS322}, issn={0883-4237}, mr={2741815}}
\end{barticle}
\bptok{imsref}\vadjust{\vspace*{-6pt}\goodbreak}%
\endbibitem

\bibitem[\protect\citeauthoryear{Patterson, Hannig and Iyer}{2004}]{PattersonHannigIyer2004}
\begin{bmisc}[auto:STB|2014/05/27|07:42:02]
\bauthor{\bsnm{Patterson},~\bfnm{P.}\binits{P.}},
\bauthor{\bsnm{Hannig},~\bfnm{J.}\binits{J.}} \AND
\bauthor{\bsnm{Iyer},~\bfnm{H.~K.}\binits{H.~K.}}
(\byear{2004}).
\bhowpublished{Fiducial generalized confidence intervals for proportion of conformance.
Technical Report
2004/11,
Colorado State Univ., Fort Collins, CO.}
\end{bmisc}
\bptok{imsref}%
\endbibitem

\bibitem[\protect\citeauthoryear{Salome}{1998}]{Salome1998}
\begin{bmisc}[auto:STB|2014/05/27|07:42:02]
\bauthor{\bsnm{Salome},~\bfnm{D.}\binits{D.}}
(\byear{1998}).
\bhowpublished{Staristical inference via fiducial methods. Ph.D.
thesis,
Univ. Groningen}.
\end{bmisc}
\bptok{imsref}%
\endbibitem

\bibitem[\protect\citeauthoryear{Schweder and Hjort}{2002}]{SchwederHjort2002}
\begin{barticle}[mr]
\bauthor{\bsnm{Schweder},~\bfnm{Tore}\binits{T.}} \AND
\bauthor{\bsnm{Hjort},~\bfnm{Nils~Lid}\binits{N.~L.}}
(\byear{2002}).
\btitle{Confidence and likelihood}.
\bjournal{Scand. J. Stat.}
\bvolume{29}
\bpages{309--332}.
\bnote{Large structured models in applied sciences; challenges for statistics (Grimstad, 2000)}.
\bid{doi={10.1111/1467-9469.00285}, issn={0303-6898}, mr={1909788}}
\end{barticle}
\bptok{imsref}%
\endbibitem

\bibitem[\protect\citeauthoryear{Singh, Xie and Strawderman}{2005}]{SinghXieStrawderman2005}
\begin{barticle}[mr]
\bauthor{\bsnm{Singh},~\bfnm{Kesar}\binits{K.}},
\bauthor{\bsnm{Xie},~\bfnm{Minge}\binits{M.}} \AND
\bauthor{\bsnm{Strawderman},~\bfnm{William~E.}\binits{W.~E.}}
(\byear{2005}).
\btitle{Combining information from independent sources through confidence distributions}.
\bjournal{Ann. Statist.}
\bvolume{33}
\bpages{159--183}.
\bid{doi={10.1214/009053604000001084}, issn={0090-5364}, mr={2157800}}
\end{barticle}
\bptok{imsref}%
\endbibitem

\bibitem[\protect\citeauthoryear{Sonderegger and Hannig}{2014}]{SondereggerHannig2013}
\begin{bincollection}[auto:STB|2014/05/27|07:42:02]
\bauthor{\bsnm{Sonderegger},~\bfnm{D.}\binits{D.}} \AND
\bauthor{\bsnm{Hannig},~\bfnm{J.}\binits{J.}}
(\byear{2014}).
\btitle{Fiducial theory for free-knot splines}.
In \bbooktitle{Contemporaly Developments in Statistical Theory, a Festschrift in Honor of Professor Hira L. Koul}
(\beditor{\bfnm{T.~N.}\binits{T.~N.}~\bsnm{Sriraus}}, ed.)
\bpages{155--189}.
\bpublisher{Springer},\break 
\blocation{Berlin}.
\end{bincollection}
\bptok{imsref}%
\endbibitem

\bibitem[\protect\citeauthoryear{Stevens}{1950}]{Stevens1950}
\begin{barticle}[mr]
\bauthor{\bsnm{Stevens},~\bfnm{W.~L.}\binits{W.~L.}}
(\byear{1950}).
\btitle{Fiducial limits of the parameter of a discontinuous distribution}.
\bjournal{Biometrika}
\bvolume{37}
\bpages{117--129}.
\bid{issn={0006-3444}, mr={0035955}}
\end{barticle}
\bptok{imsref}%
\endbibitem

\bibitem[\protect\citeauthoryear{Taraldsen and Lindqvist}{2013}]{TaraldsenLindqvist2013}
\begin{barticle}[mr]
\bauthor{\bsnm{Taraldsen},~\bfnm{Gunnar}\binits{G.}} \AND
\bauthor{\bsnm{Lindqvist},~\bfnm{Bo~Henry}\binits{B.~H.}}
(\byear{2013}).
\btitle{Fiducial theory and optimal inference}.
\bjournal{Ann. Statist.}
\bvolume{41}
\bpages{323--341}.
\bid{doi={10.1214/13-AOS1083}, issn={0090-5364}, mr={3059420}}
\end{barticle}
\bptok{imsref}%
\endbibitem

\bibitem[\protect\citeauthoryear{Tsui and Weerahandi}{1989}]{TsuiWeerahandi1989}
\begin{barticle}[mr]
\bauthor{\bsnm{Tsui},~\bfnm{Kam-Wah}\binits{K.-W.}} \AND
\bauthor{\bsnm{Weerahandi},~\bfnm{Samaradasa}\binits{S.}}
(\byear{1989}).
\btitle{Generalized {$p$}-values in significance testing of hypotheses in the presence of nuisance parameters}.
\bjournal{J. Amer. Statist. Assoc.}
\bvolume{84}
\bpages{602--607}.
\bid{issn={0162-1459}, mr={1010352}}
\end{barticle}
\bptok{imsref}%
\endbibitem

\bibitem[\protect\citeauthoryear{Tsui and Weerahandi}{1991}]{TsuiWeerahandi1991}
\begin{barticle}[mr]
\bauthor{\bsnm{Tsui},~\bfnm{Kam-Wah}\binits{K.-W.}} \AND
\bauthor{\bsnm{Weerahandi},~\bfnm{Samaradasa}\binits{S.}}
(\byear{1991}).
\btitle{Corrections: ``Generalized $p$-values in significance testing
of hypotheses in the presence of nuisance parameters''}.
\bnote{[\textit{J. Amer.
Statist. Assoc.} \textbf{84} (1989)  602--607. MR1010352]}.
\bjournal{J. Amer. Statist. Assoc.}
\bvolume{86}
\bpages{256}.
\bptnote{check related}%
\bid{issn={0162-1459}, mr={1137115}}
\end{barticle}
\bptok{imsref}%
\endbibitem

\bibitem[\protect\citeauthoryear{Wang, Hannig and Iyer}{2012}]{WangHannigIyer2012}
\begin{barticle}[mr]
\bauthor{\bsnm{Wang},~\bfnm{C.~M.}\binits{C.~M.}},
\bauthor{\bsnm{Hannig},~\bfnm{Jan}\binits{J.}} \AND
\bauthor{\bsnm{Iyer},~\bfnm{Hari~K.}\binits{H.~K.}}
(\byear{2012}).
\btitle{Fiducial prediction intervals}.
\bjournal{J. Statist. Plann. Inference}
\bvolume{142}
\bpages{1980--1990}.
\bid{doi={10.1016/j.jspi.2012.02.021}, issn={0378-3758}, mr={2903406}}
\end{barticle}
\bptok{imsref}%
\endbibitem

\bibitem[\protect\citeauthoryear{Weerahandi}{1993}]{Weerahandi1993}
\begin{barticle}[mr]
\bauthor{\bsnm{Weerahandi},~\bfnm{Samaradasa}\binits{S.}}
(\byear{1993}).
\btitle{Generalized confidence intervals}.
\bjournal{J.~Amer. Statist. Assoc.}
\bvolume{88}
\bpages{899--905}.
\bid{issn={0162-1459}, mr={1242940}}
\end{barticle}
\bptok{imsref}%
\endbibitem

\bibitem[\protect\citeauthoryear{Weerahandi}{1994}]{Weerahandi1994}
\begin{barticle}[mr]
\bauthor{\bsnm{Weerahandi},~\bfnm{Samaradasa}\binits{S.}}
(\byear{1994}).
\btitle{Correction: ``Generalized confidence intervals''}.
\bnote{[\textit{J. Amer.
Statist. Assoc.} \textbf{88} (1993)  899--905. MR1242940]}.
\bjournal{J. Amer. Statist. Assoc.}
\bvolume{89}
\bpages{726}.
\bid{issn={0162-1459}, mr={1294096}}
\bptnote{check related}%
\end{barticle}
\bptok{imsref}%
\endbibitem

\bibitem[\protect\citeauthoryear{Weerahandi}{1995}]{Weerahandi1995}
\begin{bbook}[mr]
\bauthor{\bsnm{Weerahandi},~\bfnm{Samaradasa}\binits{S.}}
(\byear{1995}).
\btitle{Exact Statistical Methods for Data Analysis}.
\bseries{Springer Series in Statistics}.
\bpublisher{Springer},
\blocation{New York}.
\bid{doi={10.1007/978-1-4612-0825-9}, mr={1316663}}
\end{bbook}
\bptok{imsref}%
\endbibitem

\bibitem[\protect\citeauthoryear{Wilkinson}{1977}]{Wilkinson1977}
\begin{barticle}[mr]
\bauthor{\bsnm{Wilkinson},~\bfnm{G.~N.}\binits{G.~N.}}
(\byear{1977}).
\btitle{On resolving the controversy in statistical inference}.
\bjournal{J. R. Stat. Soc. Ser. B Stat. Methodol.}
\bvolume{39}
\bpages{119--171}.
\bid{issn={0035-9246}, mr={0652326}}
\bptnote{check related}%
\end{barticle}
\bptok{imsref}%
\endbibitem

\bibitem[\protect\citeauthoryear{Xie and Singh}{2013}]{XieSingh2013}
\begin{barticle}[mr]
\bauthor{\bsnm{Xie},~\bfnm{Min-ge}\binits{M.-g.}} \AND
\bauthor{\bsnm{Singh},~\bfnm{Kesar}\binits{K.}}
(\byear{2013}).
\btitle{Confidence distribution, the frequentist distribution estimator of a parameter: A~review}.
\bjournal{Internat. Statist. Rev.}
\bvolume{81}
\bpages{3--39}.
\bid{doi={10.1111/insr.12000}, issn={0306-7734}, mr={3047496}}
\end{barticle}
\bptok{imsref}%
\endbibitem

\bibitem[\protect\citeauthoryear{Xie, Singh and Strawderman}{2011}]{XieSinghStrawderman2011}
\begin{barticle}[mr]
\bauthor{\bsnm{Xie},~\bfnm{Minge}\binits{M.}},
\bauthor{\bsnm{Singh},~\bfnm{Kesar}\binits{K.}} \AND
\bauthor{\bsnm{Strawderman},~\bfnm{William~E.}\binits{W.~E.}}
(\byear{2011}).
\btitle{Confidence distributions and a unifying framework for meta-analysis}.
\bjournal{J. Amer. Statist. Assoc.}
\bvolume{106}
\bpages{320--333}.
\bid{doi={10.1198/jasa.2011.tm09803}, issn={0162-1459}, mr={2816724}}
\end{barticle}
\bptok{imsref}%
\endbibitem

\bibitem[\protect\citeauthoryear{Xie et~al.}{2013}]{XieEtAl2013}
\begin{barticle}[mr]
\bauthor{\bsnm{Xie},~\bfnm{Minge}\binits{M.}},
\bauthor{\bsnm{Liu},~\bfnm{Regina~Y.}\binits{R.~Y.}},
\bauthor{\bsnm{Damaraju},~\bfnm{C.~V.}\binits{C.~V.}} \AND
\bauthor{\bsnm{Olson},~\bfnm{William~H.}\binits{W.~H.}}
(\byear{2013}).
\btitle{Incorporating external information in analyses of clinical trials with binary outcomes}.
\bjournal{Ann. Appl. Stat.}
\bvolume{7}
\bpages{342--368}.
\bid{doi={10.1214/12-AOAS585}, issn={1932-6157}, mr={3086422}}
\end{barticle}
\bptok{imsref}%
\endbibitem

\bibitem[\protect\citeauthoryear{Zhang and Liu}{2011}]{ZhangLiu2011}
\begin{barticle}[mr]
\bauthor{\bsnm{Zhang},~\bfnm{Jianchun}\binits{J.}} \AND
\bauthor{\bsnm{Liu},~\bfnm{Chuanhai}\binits{C.}}
(\byear{2011}).
\btitle{Dempster--{S}hafer inference with weak beliefs}.
\bjournal{Statist. Sinica}
\bvolume{21}
\bpages{475--494}.
\bid{doi={10.5705/ss.2011.022a}, issn={1017-0405}, mr={2829843}}
\end{barticle}
\bptok{imsref}\vspace*{-6pt}
\endbibitem

\end{thebibliography}
\end{document}